# Discovery of self-assembled Ru/Si heterostructures with unique periodic nanostripe patterns boosting hydrogen evolution

Weizheng Cai[1†], Xinyi He[2†], Tian-Nan Ye[3], Xinmeng Hu[1], Chuanlong Liu[1], Masato Sasase[2], Masaaki Kitano[2], Toshio Kamiya[2], Hideo Hosono[2*], Jiazhen Wu[1, 4 *]

[1]Shenzhen Key Laboratory of Micro/Nano-Porous Functional Materials, Department of Materials Science and Engineering, Southern University of Science and Technology, Shenzhen, 518055, China.

[2]MDX Research Center for Element Strategy, International Research Frontiers Initiative, Tokyo Institute of Technology, Yokohama, 226-8503, Japan.

[3]Frontiers Science Center for Transformative Molecules, School of Chemistry and Chemical Engineering, Shanghai Jiao Tong University, Shanghai, 200240, China.

[4]Guangdong Provincial Key Laboratory of Functional Oxide Materials and Devices, Southern University of Science and Technology, Shenzhen, 518055, China.

*Corresponding author. E-mail: hosono@msl.titech.ac.jp, wujz@sustech.edu.cn

† These authors contributed equally to this work.

**Two-dimensional (2D) heterostructuring is a versatile methodology for designing nanoarchitecture catalytic systems that allow for reconstruction and modulation of interfaces and electronic structures. However, catalysts with such structures are extremely scarce due to limited synthetic strategies. Here, we report a highly ordered 2D Ru/Si nano-heterostructures (RSHS) by acid etching of the LaRuSi electride. RSHS shows a superior electrocatalytic activity for hydrogen evolution with an overpotential of 14 mV at 10 mA/cm$^2$ in alkaline media. Both experimental analysis and first-principles calculations demonstrate that the electronic states of Ru can be tuned by strong interactions of the interfacial Ru-Si, leading to an optimized hydrogen adsorption energy. Moreover, due to the synergistic effect of Ru and Si, the energy barrier of water dissociation is significantly reduced. The unique nanostripe structure with abundant interfaces in RSHS will provide a paradigm for construction of efficient catalysts with tunable electronic states and dual active sites.**

**Introduction**

A heterostructure is a hybrid assembly composed of different solid-state materials forming interfaces. It not only inherits the properties of the component materials but also possesses unique reconstructed lattice and electronic structures at the interfaces, and therefore has attracted a surge of interests in the fields of electronics, spintronics, photonics, energy storage and catalysis.[1-6] In the area of heterogeneous catalysis, constructing a heterostructure has been regarded as an effective method for tunning and optimization of the electronic structure and catalytic performance of the active sites.[4-10] Such an approach has been successfully applied for various catalytic reactions [5, 6, 11-13], including photocatalytic $CO_2$ reduction[14], electrocatalytic hydrogen evolution reaction (HER),[15, 16] etc.

Typically, heterostructures are fabricated by layer-by-layer physical vapor deposition techniques such as molecular beam epitaxy, pulsed laser deposition and sputtering, which are, however, inefficient for catalytic applications. Recently, it has been reported that, superlattice-like heterostructure catalysts could be readily prepared by utilizing electrostatic interactions between oppositely charged nanosheets[17, 18]. For instance, electropositive $Ni_{2/3}Fe_{1/3}$ LDH nanosheets and electronegative graphene oxide (GO) were shown to be self-assembled and alternately stacked, forming a $Ni_{2/3}Fe_{1/3}$/GO heterostructure[19]. However, adjusting the surface charges of a two-dimensional (2D) nanomaterial is highly challenging, hindering the application of this method for development of new heterostructure catalysts[20, 21]. As a consequence, the number and category of reported heterostructure catalysts are very limited, and the exploration of new such materials is continuing apace.

In the present work, we propose a simple chemical etching method as an efficient approach for preparation of heterostructures. This method has been widely used for producing 2D MXenes from MAX phases (M denotes an early transition element, A is a p-block element, X is nitrogen and/or carbon).[22, 23] Due to the distinctly different bond strength of M-A (weak) and M-X (strong), A atoms can be selectively removed from MAX structure, leading to the formation of the $M_{n+1}X_nT_x$ (T is the functional group of O, F, OH, etc., $x$ is a variable and n = 1, 2, and 3) nanosheets. $Ti_2CT_x$ and $Ti_3C_2T_x$ have been frequently reported as typical examples of MXenes[24, 25]. Another noteworthy example is Raney nickel, produced by selectively leaching aluminum from a nickel-aluminum alloy under controlled conditions. This process results in a porous, high-surface-area structure.[26] A similar approach has been employed for exfoliation of 2D electrides (such as $Ca_2N$, $Y_2C$ and $Sc_2C$)[27-29], in which, electrons confined in interstitial spaces serve as anions ($e^-$). For instance, the much weaker Ca-$e^-$-Ca interlayer interactions in comparison to Ca-N interactions facilitated the preparation of $Ca_2N$ monolayers via an organic-intercalation method[30]. In fact, $Ca_2N$, $Y_2C$ and $Sc_2C$ are isostructural to $Ti_2C$, and can be regarded as MXenes without functional groups[31]. However, these 2D electride materials are chemically so active that can hardly be applied for practical catalysis. Recently, chemically stable quasi-2D electrides *RE*RuSi (*RE* is a rare-earth element) have been discovered as high-efficiency catalysts for ammonia synthesis and HER[32-35]. Their Ru-Si layers are separated by and sandwiched in *RE-e⁻-RE* layers (Fig. 1a), providing a possibility to obtain 2D Ru-Si nanosheets by etching the basic *RE-e⁻-RE* using an acid. Actually, it has been demonstrated that the surface *RE* atoms could be effectively removed using the weak acidic chelating reagent EDTA.[32, 35]

Herein, inspired by the reported work of MXenes and 2D electrides, we employ a strong acid (1M hydrochloric acid) as the etchant, and successfully remove the La-$e^-$-La layers from the LaRuSi electride (LRS). The residual Ru-Si layers are reconstructed and self-assembled, producing high-order Ru/Si nano-heterostructures (RSHS). RSHS exhibits a high surface area of ~235 m$^2$/g and abundant Ru/Si interfaces with gradually varied concentrations of Ru and Si across the interface. The unique structure enables effective adsorptions of H and OH on Ru and its neighboring Si sites, respectively, leading to accelerated kinetics of water dissociation. Therefore, the RSHS catalyst exhibits a remarkable HER performance with an overpotential of only 14 mV at a current density of 10 mA/cm$^2$ in an alkaline medium. The catalyst also shows a long-term stability with only a small increase in overpotential after 70 hours at 100 mA/cm$^2$.

**Results and discussion**

To prepare RSHS, the LRS precursor was firstly synthesized by an arc-melting method, as reported previously (Fig. 1a).[34] LRS is a typical intermetallic electride, which adopts a tetragonal structure with the space group P4/nmm (Fig. 1a and fig. S1). As described in the introduction section, its quasi-2D nature enables the chemical etching and exfoliation of the material. The X-ray diffraction (XRD) pattern indicates that the obtained LRS sample is of high quality with only a small quantity of LaRu$_2$Si$_2$ impurity phase (Fig. 2a and fig. S2). As shown in figs. S3 and S4, the hand-milled particle exhibits a large particle size (> 1 $\mu$m) and a homogeneous element distribution of La (35.6%), Ru (33.4%) and Si (31%).

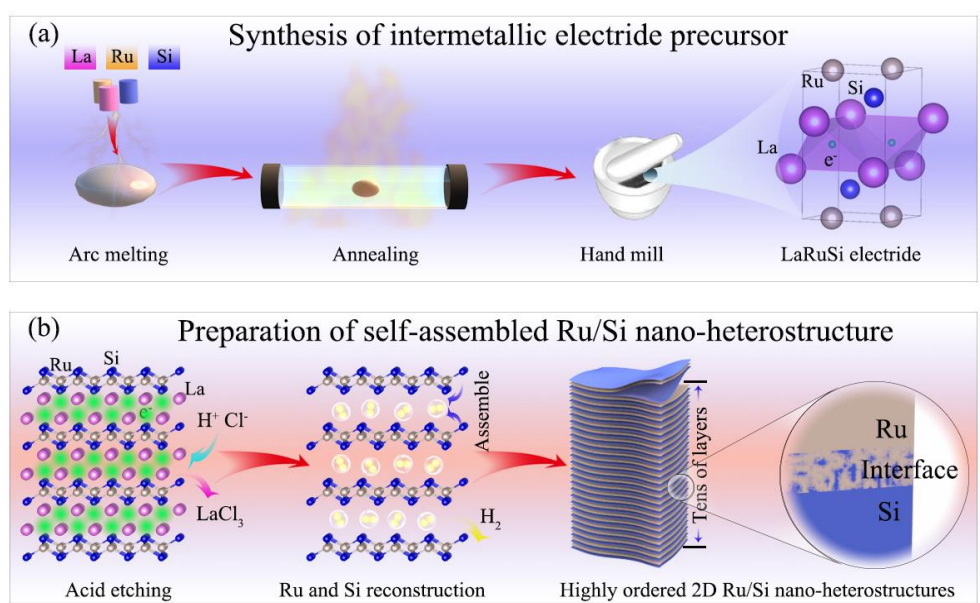

**Fig. 1**. Schematics of the materials synthetic strategy. (a) The preparation of LaRuSi electride precursor. (b) The preparation of RSHS by acid etching.

The RSHS sample was subsequently obtained by acid etching of LRS with 1 M HCl (Fig. 1b). During the acid etching experiment, we observed the emergence of noticeable bubbles, which were detectable by a portable hydrogen leak detector. After chemical etching, the X-ray diffraction peaks arising from the parent phase LRS disappeared (Fig. 2a), indicating that it is completely corroded. Instead, very broad diffraction peaks corresponding to the Ru metal and the uncorroded impurity phase $LaRu_2Si_2$ are observed. The chemical composition of RSHS was examined by energy dispersive X-ray spectroscopy (EDS) measurements (Fig. 2c, Fig. S5). The analysis results demonstrate that La is mostly removed (1.2%), likely forming $LaCl_3$ (Fig. 1b), and the ratio of Ru (51.8%) and Si (47%) remains nearly unchanged, validating our selective etching strategy. It is noted from the field emission scanning electron microscopy (FE-SEM) images that the particle size remains unaltered compared to that of LRS (Fig. 2b, Fig. S6). This result suggests that the chemical etching occurs topotactically at a nanoscale. Indeed, as the image is magnified, a nano-layered structure similar to MXenes is observed (Fig. 2d), and the surface area is largely enhanced from ~1 $m^2$/g (LRS) to ~235 $m^2$/g (RSHS).

The resulting nano structure was further characterized by high angle annular dark-field scanning transmission electron microscopy (HAADF-STEM). As shown in Fig. 2e and Fig. S7, RSHS displays a highly ordered hair-like nano-stripe structure with a period of around 5.5 nm. The periodicity can be confirmed by the equidistant bright spots along a uniaxial direction in the fast Fourier transform (FFT) image (Fig. 2e). According to the working mechanism of HAADF, bright lines with a thickness of around 2.2 nm and dark lines with a thickness of around 3.3 nm are identified with materials mainly containing Ru and Si, respectively (Fig. 2f, g). The bright-field scanning transmission electron microscopy (BF-STEM) images give similar results with bright and dark lines indicating Si and Ru, respectively (Fig. 2h and Fig. S8). The element distribution of Ru and Si is further examined by the EDS line-scan analysis and elemental mapping (Fig. 2h and Fig. S9). Clear Ru and Si stripes with gradually varied concentrations of Ru and Si across the interface are observed. To gain more specific details of the heterolayer, atomic-resolved images were measured as demonstrated in Fig. 2i and Fig. S10. In the Ru layer, clear lattice structure of Ru metal is shown, consistent with the XRD observations (Fig. 2a). In the Si layer, however, no lattice structure is presented, indicating that it is an amorphous phase. It is noteworthy that, the interface, which is composed of both Ru and Si (Fig. 1b and 2i, Fig. S9), should also exhibit an amorphous structure.

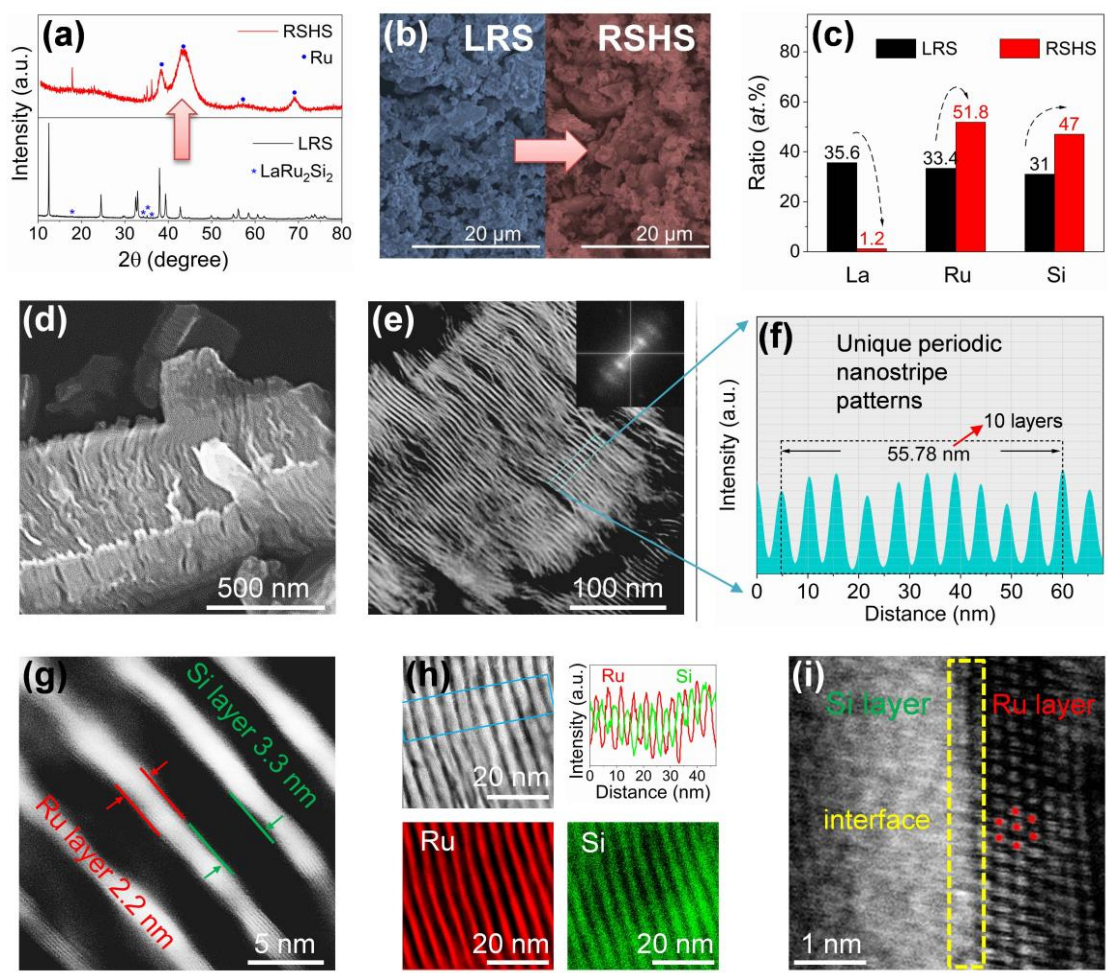

**Fig. 2.** Structure and morphology of RSHS and its precursor LRS. (**a**) XRD patterns. (**b**) FE-SEM images. (**c**) Chemical composition ratios determined by EDS. (**d**) Magnified FE-SEM image of RSHS. (**e**) HAADF-STEM image of RSHS at low magnification, the inset is the corresponding Fourier transform image. (**f**) Intensity profile for the selected area in (e). (**g**) HAADF-STEM image of RSHS at high magnification. (**h**) EDS line scan analysis and elemental mapping for RSHS. (**i**) BF-STEM image of Ru/Si interface at atomic level.

To corroborate our etching strategy using an electride, the etching experiment was also conducted on a non-electride material, LaRu$_2$Si$_2$, which has a similar crystal structure to LRS but only one single La layer (sandwiched by RuSi) instead of a La-$e^-$-La layer (Fig. S11). Different from LRS, LaRu$_2$Si$_2$ remains nearly unchanged before and after chemical etching, as indicated by the XRD (Fig. S12 and Fig. 2a), FE-SEM, and EDS (Figs. S13 and S14) measurements results. This observation suggests that the single La layer in LaRu$_2$Si$_2$ cannot be easily removed, and the double La layers with anionic electrons (La-$e^-$-La) should play an important role in the etching process of LRS.

To further investigate the etching process, a simple *quasi-operando* experiment was performed (Fig. S15), and partially etched LRS samples (intermediate state) were obtained (Fig. S16 and S17). In some areas, LRS is already transformed into RSHS as revealed by the nano-layered structure in the SEM image (Fig. S16), indicating that the chemical etching and self-assembly processes are extremely rapid. In other areas, the etching is not completed, and a broken and wrinkled surface is observed (Fig. S17e). Upon further etching with La atoms fully removed, Ru and Si are self-organized, and RSHS is formed (Fig. S17f). It is noteworthy that, different from the selective etching of A from MAX phases[23, 25], in which, $M_{n+1}X_nT_x$ monolayer can be produced, removal of La does not lead to Ru-Si monolayers. This is mainly ascribed to the weak stability of pristine RuSi in the LRS structure, which can be simply understood based on the valence states of the component elements. Previous reports have demonstrated that both Si ($-0.42|e|$) and Ru ($-0.73|e|$) in LRS are negatively charged with electrons donated from La ($+1.15|e|$)[34, 36]. Therefore, without La, the Ru-Si chemical bonding cannot be preserved due to electron deficiency, then the lattice will be reconstructed, and the Ru/Si atoms will be aggregated, forming Ru/Si nano-heterostructures. The etching process is schematically demonstrated in Fig. S17. While in MAX phases, electron transfer occurs from M to A and X. Without A, the system will become electron rich, and with the help from electronegative functional groups of T, the monolayer of $M_{n+1}X_nT_x$ can be stabilized.

To gain insight into the nano-heterostructures, the bulk chemical state was characterized by X-ray absorption near edge structure (XANES). As shown in Fig. 3a, the Ru K-edge of RSHS is located between those of a Ru foil and $RuO_2$, demonstrating that the Ru in RSHS is partially oxidized with a valence state of $Ru^{\delta+}$ ($0 < \delta < 4$). The atomic coordination environment and the local structure of Ru were analyzed by the extended X-ray absorption fine structure (EXAFS). In the k3-weighted Fourier transform (FT)-EXAFS spectra (Fig. 3b), two main peaks located at 1.52 Å and 2.35 Å are observed for RSHS, corresponding to Ru-O bonds (in $RuO_2$) and Ru-Ru bonds (in Ru metal), respectively. It should be noted that a small shoulder peak near 2.0 Å is also detected. According to the reported literatures[32-34], the appearance of this shoulder is probably due to the contributions from Ru-Si bonds. The k-weighted EXAFS data were further analyzed using the wavelet transform, and the results are displayed in Fig. 3c. There are two intensity peaks for RSHS. The first intensity maximum (k = 5.1 $Å^{-1}$), which resembles that of Ru-O in $RuO_2$ (k = 5.2 $Å^{-1}$), is assigned to the contributions from Ru-O bonds, indicating that Ru is partially oxidized upon chemical etching. The second intensity maximum (k = 8.6 $Å^{-1}$) looks similar to that of Ru-Ru in a Ru metal (k = 9.6 $Å^{-1}$), whereas its k value is slightly smaller. This suggests that, in addition to the Ru metal, which has been demonstrated by the XRD and STEM results (Fig. 2a and 2i), Ru-Si compounds should also exist in RSHS, in agreement with the EDS data at the Ru-Si interface (Fig. 2h). However, as Ru-Si and Ru-O bonds are arranged disorderedly, they could not be detected by diffraction techniques.

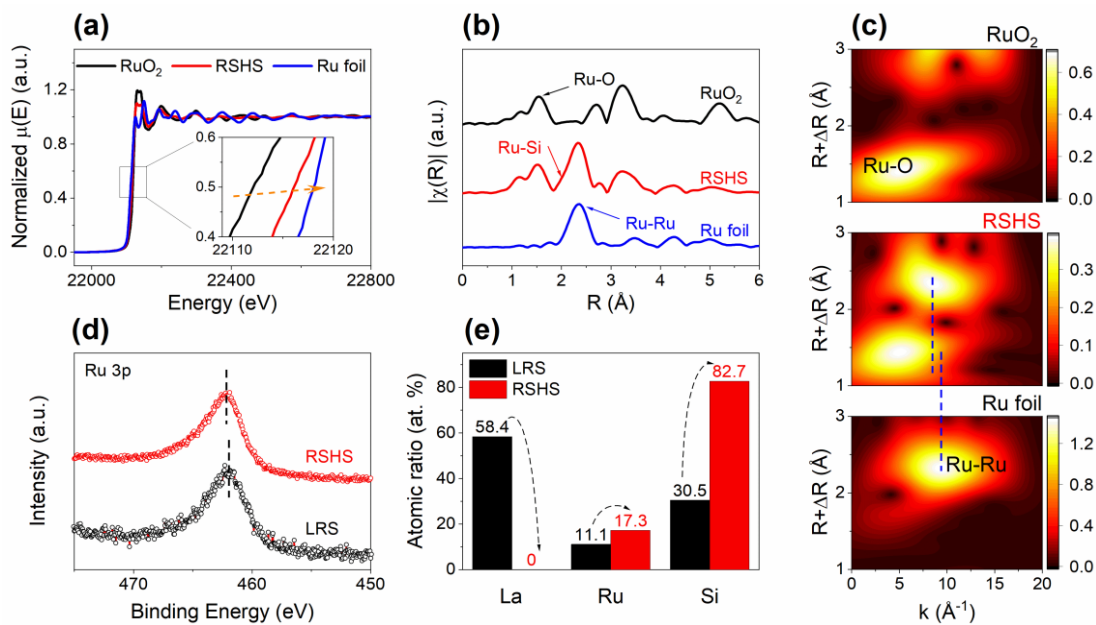

**Fig. 3**. Electronic structure and chemical information of RSHS. (**a**) Ru K-edge XANES spectra. (**b**) k3-weighted Fourier transformed EXAFS data. (**c**) Wavelet transform of the k-weighted EXAFS data. (**d**) High-resolution XPS spectra of Ru 3p. (**e**) Surface element ratios of La, Ru, Si deduced from XPS analysis.

The surface chemical information of RSHS was examined by high-resolution X-ray photoelectron spectroscopy (XPS) measurements. As displayed in Fig. 3d and Fig. S18, all the spectra, including those of Ru 3p, Si 2s and O 1s, exhibit positive shifts after acid etching, manifesting that both surficial Ru and Si elements in RSHS are oxidized. The positive shift in O 1s from 530.9 eV to 532.4 eV suggests a change of the main oxidation species from La-O to Si-O/Ru-O.[37, 38] It is worth noting that the signals of La 3d disappear for RSHS (Fig. S18a), indicating that the surface La is completely dissolved and removed. Based on the XPS spectra, the composition ratios of La, Ru, and Si were estimated for LRS and RSHS (Fig. 3e). Upon selective etching, the content of La decreases from 58.4% to 0, and the contents of Ru and Si increase from 11.1% and 30.5% to 17.3% and 82.7%, respectively.

The as-prepared RSHS sample was employed as an electrocatalyst for HER. Its catalytic performance was evaluated in 1 M KOH using a typical three-electrode system, where the catalyst was loaded on a glassy carbon electrode. As shown in Fig. 4a, RSHS exhibits an excellent activity for hydrogen evolution, which requires an overpotential of only 14 mV to achieve 10 mA/cm$^2$, approximately 200 mV less than the pristine LRS. Its activity is also much better than those of commercial Ru/C and Pt/C catalysts, which require higher overpotentials of 72 mV and 34 mV (at 10 mA/cm$^2$), respectively, demonstrating RSHS as one of the best HER catalysts in alkaline media (see more details in Fig. 4b, Fig. S19 and Table S1).

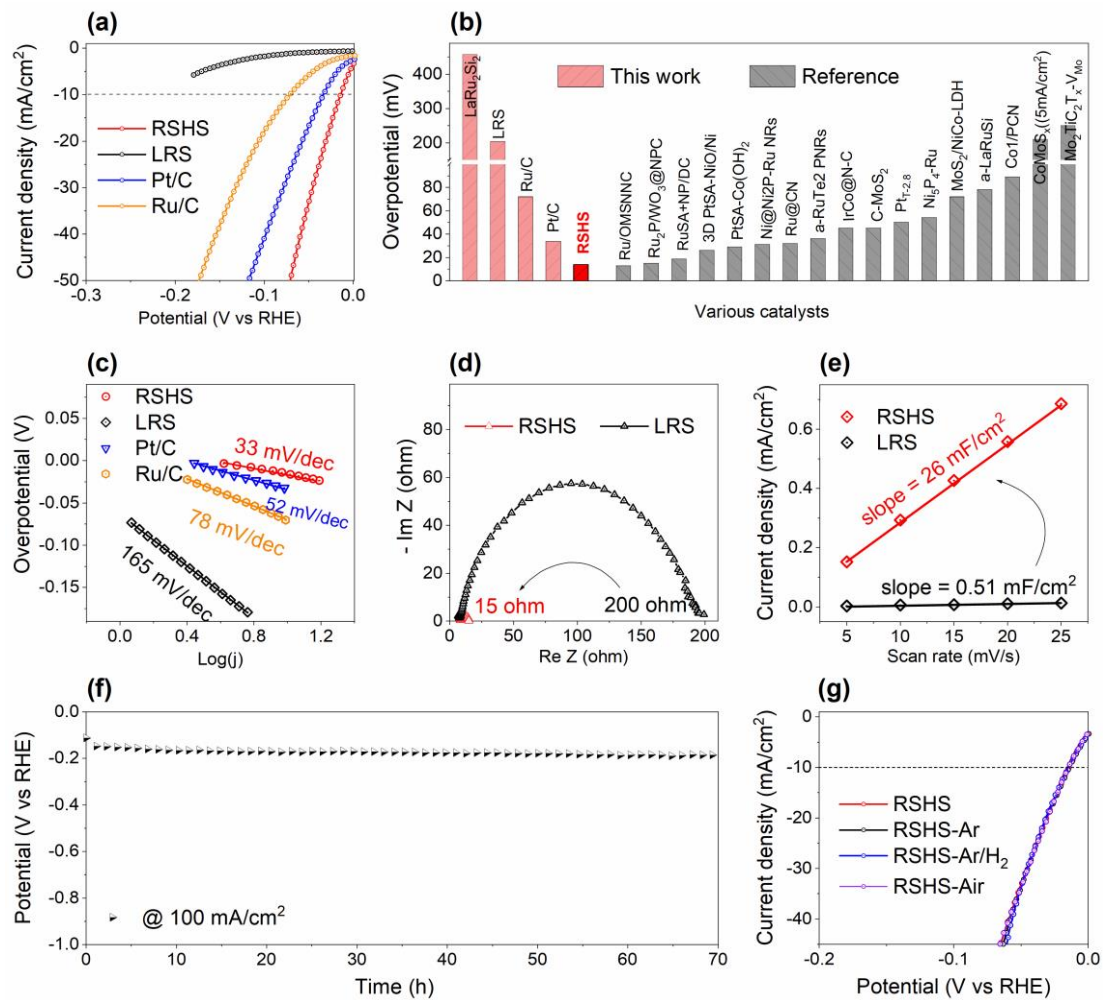

**Fig. 4**. HER performance of RSHS. (**a**) Linear sweep voltammetry (LSV) curves of as-prepared catalysts on glassy carbon for LRS, RSHS, and commercial Ru/C and Pt/C. (**b**) Overpotentials at 10 mA/cm$^2$ of various catalysts. (**c**) The Tafel curves for various catalysts. (**d**) Nyquist plots for LRS and RSHS at − 0.09 V vs RHE (reversible hydrogen electrode). (**e**) Electrochemical double-layer capacitance of LRS and RSHS. (**f**) Chronopotentiometry performance under a constant current density of 100 mA/cm$^2$ up to 70 h. (**g**) LSV curves of RSHS, RSHS-Ar, RSHS-Ar/H$_2$, RSHS-Air.

To have a deep understanding of the superior catalytic performance, the reaction kinetics was analyzed based on the Tafel slope and faradaic resistance. The Tafel slope is a measure of an electrochemical reaction rate as a relationship to the overpotential. In general, a lower Tafel slope indicates a faster kinetics and a higher catalytic efficiency. The Tafel slope curves for the investigated catalysts are plotted in Fig. 4c. Among these catalysts, RSHS exhibits a lower Tafel slope (33 mV/dec) than Pt/C (52 mV/dec), Ru/C (78 mV/dec) and LRS (165 mV/dec), manifesting it as a highly efficient HER catalyst. In addition to the Tafel slope, electrochemical impedance spectroscopy (EIS) analysis was also performed for LRS and RSHS. The faradaic resistance, which is a measure of the electrical resistance that arises from the charge transfer at an electrode surface during a reaction, can be roughly estimated from the impedance of the system. According to the *in situ* Nyquist plots (Fig. 4d), the faradaic resistance of RSHS is notably smaller than that of LRS, suggesting the presence of a much faster electron transfer efficiency in RSHS.

Moreover, the electrochemical active surface areas (ECSA) of LRS and RSHS were estimated based on the double-layer capacitance ($C_{dl}$), which can be obtained by the cyclic voltammetry at different scan rates (Fig. S20). As expected, RSHS has a much larger ECSA ($C_{dl}$ = 26 mF/cm$^2$) than LRS ($C_{dl}$ = 0.51 mF/cm$^2$) (Fig. 4e), consistent with the Brunauer–Emmett–Teller (BET) analysis results for RSHS (~235 m$^2$/g) and LRS (~1 m$^2$/g). The catalytic durability of RSHS was examined by loading the catalyst over a three-dimensional (3D) porous nickel foam (thickness: 0.08 cm). As shown in Fig. 4f, the overpotential at 100 mA/cm$^2$ only exhibits a small increase after 70 h of continuous reaction, revealing a robust catalytic stability and great application potentials.

Uncovering the catalytic active species is an intriguing issue that awaits to solve for RSHS. Based on the XRD, STEM and EXAFS measurements results (Fig.2 and 3), RSHS was shown to consist of crystalline Ru metal and amorphous Ru-O, Si-O, and Ru-Si compounds. The distinct overpotentials and Tafel slopes of RSHS and Ru/C (Fig. 4b, c) demonstrate that the excellent catalytic performance of RSHS is not due to the crystalline Ru metal. According to literatures[39, 40], $RuO_2$ is also not an efficient catalyst for alkaline HER with an overpotential of 80 mV (at 10 mA/cm$^2$) and a Tafel slope of more than 100 mV/dec, which exclude it as an effective component in RSHS. Furthermore, the overpotential of residual $LaRu_2Si_2$ is around 458 mV (Fig. S21), indicating that it cannot be attributed to the high activity. Therefore, the amorphous Ru-Si compound at the interface is considered as the potent constituent of RSHS that possesses the high catalytic activity.

To confirm the Ru-Si active sites, control experiments were performed on basis of various RSHS catalysts: i.e. RSHS-Air, RSHS-Ar, and RSHS-Ar/H$_2$, which were prepared by annealing RSHS samples at 200 °C for 80 min in air, Ar, and 95% Ar - 5% H$_2$ atmosphere, respectively. Interestingly, these materials show similar LSV curves and Tafel slopes (Fig. 4g, Fig. S22), even though their XRD patterns are notably different (Fig. S23). After heat treatment, RSHS-Air is mostly oxidized as indicated by the broad diffraction peaks of $RuO_2$, RSHS-Ar/H$_2$ is partially reduced because of a large increase in the diffraction intensity of Ru, while RSHS-Ar remains practically unchanged. Further investigations indicate that the ECSAs of these catalysts are similar and independent of treatment conditions (Fig. S24). These results suggest that neither Ru metal nor Ru oxides play a major role in the RSHS's activity, supporting our idea that Ru-Si is the active site. In addition, XPS measurements were conducted to reveal the surface chemical information of these control samples (Fig. S25). The data of $Ru_4Si_3$ is also presented for comparison, as it will be shown later that the interfacial Ru/Si amorphous compound most probably adopts the $Ru_4Si_3$ structure. According to the XPS data, the valence states of Si and O only exhibit small variations upon heat treatment, while the valence state of Ru experiences a negative shift in reducing atmosphere

(Ar/H$_2$) and a positive shift in oxidizing atmosphere (air), in agreement with the XRD data. To make it clearer, a peak-differentiation-imitating analysis was performed for the XPS data of Ru 3p (Fig. S26), and four types of Ru were considered: Ru in Ru$_4$Si$_3$, Ru metal, RuO$_2$ and a satellite peak. Indeed, as with XRD results, the Ru component in RSHS-Ar shows minimal changes, and RSHS-Ar/H$_2$ (RSHS-Air) display a significant increase (decrease) in the peak ratio of Ru/RuO$_2$[41]. It should be noted that the Ru component corresponding to Ru$_4$Si$_3$ is similar for all the investigated RSHS samples, suggesting that the oxidation/reduction primarily occurs on the Ru metal/Ru-O compound rather than the Ru-Si interface. Given that Ru-Si is the active site, the similar Ru-Si content also explains the similar catalytic performance of these catalysts (Fig. S20). To further underscore the contribution of Si, we eliminated the Si species through alkali etching in 20 *wt.* % KOH at 70 °C for 5 hours. The obtain catalyst (99% Ru, 1% Si by EDS) displays an increased overpotential of 36 mV at 10 mA/cm$^2$ (Fig. S27), due to the diminished Ru-Si interface.

To elucidate the catalytic mechanism over the Ru-Si interface, it is necessary to obtain the local crystal structure of the amorphous phase. As high-temperature annealing could promote the crystallization of an amorphous material, a heat treatment was performed for RSHS in Ar atmosphere at 1200 °C for 12 hours. As expected, in addition to Ru metal, crystalline Ru$_4$Si$_3$ and SiO$_2$ are also detected after annealing (Fig. S28). This implies that the local structure of the amorphous Ru-Si is similar to that of Ru$_4$Si$_3$, because similar chemical bonding configurations could lower the crystallization energy barrier. Therefore, the crystal structure of Ru$_4$Si$_3$ was employed for theoretical simulations of the surface chemical reaction process over RSHS. The Ru$_4$Si$_3$ (010) surface, which possesses the lowest surface energy among the low-index surfaces (Fig. 5a, Table S2), was selected for investigation. The most stable adsorption configuration of H$_2$O over Ru$_4$Si$_3$ (010) was identified by calculating the adsorption energies on various Ru sites (Fig. 5b, Fig. S29). To simulate the Ru-Si interface with gradually varied concentrations of Ru and Si (Fig. 2i), two additional surfaces were constructed: i.e. Ru$_4$Si$_3$ (010) with Si partially removed and Ru$_4$Si$_3$ (010) with Ru partially removed (Fig. 5c). For comparison, Ru (001) was also studied.

With the surface models of Ru$_4$Si$_3$ (010) and Ru (001), the HER catalytic activity was evaluated by the density functional theory (DFT) calculations. The H adsorption energy ($\Delta G_\text{H}$) was computed for various surface models. It is a parameter closely associated to the associative formation of H$_2$ from the adsorbed H atoms, rendering it a crucial indicator of the HER catalytic performance. As illustrated in Fig. 5d, the Ru$_4$Si$_3$ (010) surfaces exhibit significantly reduced H adsorption strengths ($\Delta G_\text{H} = -0.147$ eV, $-0.152$ eV, and $-0.190$ eV) compared to the Ru (001) surface ($\Delta G_\text{H} = -0.562$ eV), implying a faster formation rate of H$_2$ from H* on the Ru$_4$Si$_3$ catalysts. Among the constructed Ru$_4$Si$_3$ (010) surfaces, the one with Ru partially removed exhibits the best performance (Fig. 5d). This indicates that the activity may vary with the composition ratio of Ru/Si.

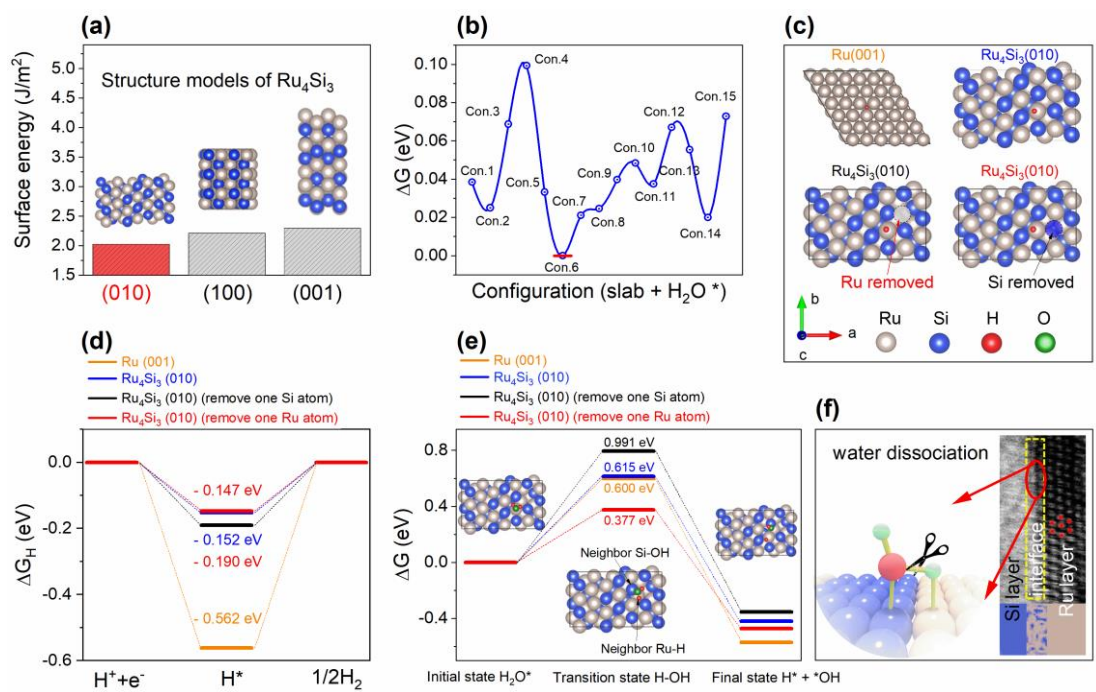

**Fig. 5**. Theoretical investigation. (**a**) Surface energies of $Ru_4Si_3$, which was used to approximate the amorphous interface in RSHS. (**b**) $H_2O$ adsorption free energy of different adsorption configurations on the Ru sites of the $Ru_4Si_3$ (010) surface. (**c**) Adsorption configurations of H on Ru sites of Ru (001), $Ru_4Si_3$ (010), $Ru_4Si_3$ (010) with one Si atom removed, and $Ru_4Si_3$ (010) with one Ru atom removed, depicted using VESTA[42] software. (**d**) Hydrogen adsorption energy profiles for different surface models. (**e**) Kinetic barriers of water dissociation for different surface models. The optimized adsorption configurations of $Ru_4Si_3$ (010) with one Ru atom removed are shown as insets. (**f**) Schematic illustration of water dissociation over the Ru-Si interface.

Furthermore, the reaction barrier ($\Delta G$) of water dissociation, which is usually a sluggish step in alkaline HER, was also evaluated (Fig. 5e and Figs. S30 to S33). Interestingly, the $\Delta G$ of water dissociation for $Ru_4Si_3$ (010) surface gradually decreases as the composition ration of Ru/Si is reduced. The pristine $Ru_4Si_3$ (010) surface exhibits a similar barrier (0.615 eV) to the Ru (001) surface (0.599 eV), whereas the $Ru_4Si_3$ (010) surface with Ru(Si) partially removed presents a significantly lower(higher) barrier of 0.376 eV(0.794 eV), demonstrating the critical role of Ru concentration in determining the HER activity. With a gradually varied Ru concentration at the Ru-Si interface of RSHS (Fig. 2i), an optimized configuration could be achieved, leading to significantly enhanced alkaline HER activity compared to the Ru/C catalyst.

Further investigation reveals that water dissociation on the $Ru_4Si_3$ (010) surface occurs via the formation of Si-OH and Ru-H (Fig. 5f, Fig. S34), which therefore can be regarded as dual active sites, analogous to transition metal/transition metal hydroxide composite catalysts.[16, 43-45] The synergistic catalytic effect of Ru and Si can be further enhanced by partially removing the surface Ru atoms, leading to a large reduction of the kinetic barrier (Fig. S32). The variation of catalytic activity with the Ru concentration is probably attributed to the modulation of the surface electronic states. Given that the electronegativity of Ru is greater than that of Si, removal of Si would lead to an electron-deficient surface, a weaker adsorption, and an increased energy barrier. While, in contrast, removal of Ru would lead to an electron-rich surface, a stronger adsorption, and a smaller energy barrier. It is noteworthy that constructing an accurate model to simulate the amorphous Ru-Si interface of RSHS is a challenging task, and the $Ru_4Si_3$ structure might not be completely accurate for this simulation. However, the present model, developed based on experimental observations, can offer a reasonable interpretation of the excellent HER activity.

**Conclusions**

In summary, we reported a highly ordered self-assembled 2D Ru/Si nano-heterostructures (RSHS), which was prepared via a simple selective chemical etching method. The 2D electride features of LRS were shown to be crucial for the acid etching and subsequent formation of the nano-heterostructures. With a high surface area and abundant Ru-Si interfaces, RSHS exhibited a potent catalytic activity in alkaline HER with an overpotential of only 14 mV (10 mA/cm$^2$) and a Tafel slope of only 33 mV/dec. The catalyst also exhibited a robust catalytic stability and great application potentials. By combining structure analysis and DFT calculations, we demonstrated that the excellent catalytic performance was primarily attributed to the unique Ru-Si interface with gradually varied distribution of Ru and Si. It could not only enable a synergistic catalytic effect of Ru and Si via the formation of Si-OH and Ru-H, leading to a reduced kinetic barrier for water dissociation, but also adsorb H suitably, thereby promoting the associative formation of H$_2$ from adsorbed H atoms. The present findings offer a novel strategy for the design and preparation of 2D nano-heterostructures and related catalysts.

**Methods**

**Chemicals.** All chemicals were purchased and used without further purification. Potassium hydroxide (semiconductor grade, 99.99% trace metals basis) and Nafion™ 117 solution (around 5% in a mixture of lower aliphatic alcohols and water) were purchased from Sigma-Aldrich. La (99.9%) and Si (99.99%) were purchased from Kojundo Chemical Laboratory Co., Ltd., Ru (99.9%) was purchased from Rare Metallic Co., Ltd.

**Materials preparation.** LRS, LaRu$_2$Si$_2$ samples were prepared using an arc-melting method with stoichiometric amounts of lanthanum, ruthenium, and silicon ingots under an argon atmosphere. To ensure homogeneity, the melting process was repeated 5 times for each sample, and the weight loss after the melting process was less than 0.1%. The resulting ingots had a silver color and were ground into powder using an agate motor in an Ar-filled glove box. Note that the as-melted LRS was not a single phase, and an annealing process was employed at 1000 °C for 10 days to eliminate the impurity phases.

RSHS was prepared by chemically etching La from LaRuSi using hydrochloric acid. Firstly, LaRuSi powder (around 0.2 g) was placed in a centrifugal tube, and 10 ml of 1 M HCl was slowly added. After stirring for 24 h, the mixture was centrifugated, and the formed RSHS was separated, washed 3 times with distilled water and ethanol, and dried in vacuum at room temperature for 12 h. HCl-etched $LaRu_2Si_2$ was prepared using the same method. RSHS-Ar, RSHS-Ar/$H_2$ and RSHS-Air were obtained by annealing the RSHS in Ar, 95% Ar - 5% $H_2$ and air, respectively, at 200 °C for 80 min.

**Materials Characterization.** The prepared samples were examined by powder X-ray diffraction (XRD) measurements on a Bruker D8 Advance ECO diffractometer with Cu Kα radiation (λ = 1.5418 Å). Sample morphology was checked by a field emission scanning electron microscopy (FE-SEM, TESCAN MIRA3). The Brunauer–Emmett–Teller (BET) specific surface area of RSHS was determined by nitrogen adsorption-desorption isotherms at −196 ºC using an automatic gas-adsorption instrument (BELSORPmini II, MicrotracBEL). HAADF-STEM and BF-STEM images were obtained using a JEOL JEM-ARM200F atomic resolution analytical electron microscope with an operate voltage of 200 kV. Component elements were analyzed using an energy dispersive X-ray spectroscopy (EDS). Surface chemical information was analyzed by X-ray photoelectron spectroscopy (XPS) on an ESCALAB Xi+ photoelectron spectrometer (Thermo Scientific) using a monochromatic Al Kα X-ray beam (1486.6 eV). The binding energy of each spectrum was corrected according to the C 1s peak (284.8 eV). X-ray absorption spectroscopy (XAS) measurements were conducted using a synchrotron radiation ring at the NW-10A and PF-12C beamlines (Photon Factory, KEK) using Si (311) and Si (111) single-crystal monochromators at room temperature. XAFS measurements were conducted under the transmission mode.

**Electrochemical measurements.** Electrochemical measurements were performed using a CHI 760E electrochemical workstation at room temperature with a typical three-electrode system. A rotating disk glassy carbon electrode with a diameter of 5 mm and an area of 0.196 $cm^2$ (PINE) was used as the working electrode. A carbon rod and an Ag/AgCl electrode (PINE, saturated with KCl and calibrated by a standard hydrogen electrode) were used as the counter and reference electrodes, respectively. The potential was calculated by the equation:

$$E(RHE) = E(Ag/AgCl) + 0.196 + 0.059 \times pH$$

The working electrode on glassy carbon was prepared by the following steps. Firstly, the catalyst powder (5 mg) was dispersed in a mixture of deionized water (480 μL), isopropanol (480 μL) and Nafion™ 117 solution (40 μL), and ultrasonicated for 1 hour. Subsequently, the catalyst ink suspension was pipetted (10 μL) onto a pre-cleaned glassy carbon disk electrode (loading amount, 250 μg/$cm^2$). Finally, the glassy carbon electrode was dried in air at room temperature for more than 1 hour.

Prior to electrochemical data collection, the working electrodes underwent continuous potential cycling from −0.04 to 0.2 V vs RHE until reproducible voltammograms were obtained (room temperature, 1600 rpm, and a sweep rate of 100 mV/s). The solution resistance ($R_s$) was determined from the resulting Nyquist plot and was used to correct Ohmic drop using $E_c = E_m - iR_s$, where $E_c$ is the corrected potential and $E_m$ is the measured potential. To evaluate the HER activity, linear sweep voltammetry (LSV) was conducted at a scan rate of 10 mV/s in 1 M KOH.

Electrochemical impedance spectroscopy (EIS) measurements were performed using the same CHI 760E electrochemical workstation. The EIS spectra were obtained by applying various DC potentials with the frequency ranging from 0.1 Hz to 100 KHz. To ensure a linear response of the electrode, a sinusoidal potential with an amplitude of 10 mV was applied. The electrochemically active surface area (ECSA) was estimated by a double-layer capacitance method. Firstly, a double-layer charging potential region was determined from a static CV scan. The charging current $i_c$ could be calculated from the CV curves obtained at different scan rates. With the relation between $i_c$, scan rate ($v$) and the double-layer capacitance ($C_{dl}$) as $i_c = vC_{dl}$, the ECSA was then estimated according to $\text{ECSA} = \frac{C_{dl}}{C_s}$. Here, $C_s$ is the ideal specific capacitance of a smooth planar surface (set as a constant in the present study).

**Computational details.** First-principles calculations were conducted using the projector-augmented wave (PAW) method, as implemented in the Vienna Ab initio Simulation Package (VASP).[46, 47] Si (3$s$)(3$p$), Ru (4$p$)(4$d$)(5$s$), O (2$s$)(2$p$), and H (1$s$) were treated as valence states. The generalized gradient approximation (GGA) Perdew-Burke-Ernzerhof (PBE) functional[48] was employed in all the calculations. The crystal structures of Ru and Ru$_4$Si$_3$ bulks were fully optimized with the Γ-centered $k$-mesh of 0.1 Å$^{-1}$ spacing, the corresponding plane wave cut-off energy of 550 eV, and the convergence criteria of 1×10$^{-6}$ for the energy and 0.01 eV/Å for the force. The calculated lattice parameters are $a$ = 5.22 Å, $b$ = 4.03 Å, and $c$ = 17.29 Å for Ru$_4$Si$_3$, and $a = b$ = 2.72 Å, $c$ = 4.29 Å for Ru.

The surface models were built by using the $p(1 \times 2)$ supercells for Ru$_4$Si$_3$ (010) surface and the $p(6 \times 6)$ supercells for Ru (001) surface with four atomic layers and a vacuum layer of 15 Å was incorporated to avoid interactions between the neighboring slabs. The top two layers were optimized with a plane wave cut-off energy of 450 eV, and the convergence criteria of 1×10$^{-5}$ for the energy and 0.02 eV/Å for the force. The $k$-mesh was set to the Γ-centered 2 × 3 × 1 meshes for Ru$_4$Si$_3$ and 2 × 2 × 1 meshes for Ru. The DFT-D3 functional[49, 50] with PBE-PAW potentials was applied to correct the van der Waals interaction. The climbing image nudged elastic band (CI-NEB) method[51] and the dimer method[52] were applied to determine the transition states. Each transition state was verified through the vibrational frequency analysis with only one imaginary frequency.

Surface energy is defined by the following equation,

$$\gamma = \frac{E_{slab}^{unrelax} - N \cdot E_{bulk}}{2 \cdot A} + \frac{E_{slab}^{relax} - E_{slab}^{unrelax}}{A}$$

, where $E_{slab}^{unrelax}$, $E_{slab}^{relax}$, and $E_{bulk}$ are the total energies of the unrelaxed slab model, the relaxed slab model, and the bulk model, respectively. $N$ is the number of bulk units in the slab model, and $A$ is the surface area of the slab model.

The hydrogen adsorption energy ($\Delta G_H$) is determined by the following equation,

$$\Delta G_H = E_{(H^*/slab)} - \frac{1}{2} E_{H_2} - E_{slab}$$

where $E_{(H^*/slab)}$, $E_{H_2}$, and $E_{slab}$ are the total energies of the surface slab with an adsorbed H atom, a gaseous-phase $H_2$ molecule, and a clean surface, respectively. The reaction barrier ($\Delta G$) of water dissociation was obtained by $\Delta G = E(TS) - E(IS)$, where $E(IS)$ and $E(TS)$ are the energies of the initial state (IS) and the transition state (TS), respectively. The zero-point energy corrections were added to all the above energies with the assistance of the VASPKIT code.[53]

**Reference**
1. Geim, A. K.; Grigorieva, I. V., Van der Waals heterostructures. *Nature* **2013,** *499* (7459), 419-425.
2. Liu, Y.; Weiss, N. O.; Duan, X. D.; Cheng, H. C.; Huang, Y.; Duan, X. F., Van der Waals heterostructures and devices. *Nat. Rev. Mater.* **2016,** *1* (9), 16042.
3. Pomerantseva, E.; Gogotsi, Y., Two-dimensional heterostructures for energy storage. *Nat. Energy* **2017,** *2* (7), 17089.
4. Zhao, G. Q.; Rui, K.; Dou, S. X.; Sun, W. P., Heterostructures for electrochemical hydrogen evolution reaction: A review. *Adv. Funct. Mater.* **2018,** *28* (43), 1803291.
5. Deng, D. H.; Novoselov, K. S.; Fu, Q.; Zheng, N. F.; Tian, Z. Q.; Bao, X. H., Catalysis with two-dimensional materials and their heterostructures. *Nat. Nanotechnol.* **2016,** *11* (3), 218-230.
6. Ahsan, M. A.; He, T. W.; Noveron, J. C.; Reuter, K.; Puente-Santiago, A. R.; Luque, R., Low-dimensional heterostructures for advanced electrocatalysis: an experimental and computational perspective. *Chem. Soc. Rev.* **2022,** *51* (3), 812-828.
7. Chen, P. C.; Liu, M. H.; Du, J. S. S.; Meckes, B.; Wang, S. Z.; Lin, H. X.; Dravid, V. P.; Wolverton, C.; Mirkin, C. A., Interface and heterostructure design in polyelemental nanoparticles. *Science* **2019,** *363* (6430), 959-964.
8. Prabhu, P.; Jose, V.; Lee, J. M., Heterostructured catalysts for electrocatalytic and photocatalytic carbon dioxide reduction. *Adv. Funct. Mater.* **2020,** *30* (24), 1910768.
9. Shifa, T. A.; Wang, F. M.; Liu, Y.; He, J., Heterostructures based on 2D materials: A versatile platform for efficient catalysis. *Adv. Mater.* **2019,** *31* (45), 1804828.
10. Su, J.; Li, G. D.; Li, X. H.; Chen, J. S., 2D/2D Heterojunctions for catalysis. *Adv. Sci.* **2019,** *6* (7), 1801702.
11. Green, I. X.; Tang, W.; Neurock, M.; Yates, J. T., Spectroscopic observation of dual catalytic sites during oxidation of CO on a Au/TiO2 catalyst. *Science* **2011,** *333* (6043), 736-739.
12. Li, C. W.; Ciston, J.; Kanan, M. W., Electroreduction of carbon monoxide to liquid fuel on oxide-derived nanocrystalline copper. *Nature* **2014,** *508* (7497), 504-507.
13. Schreier, M.; Héroguel, F.; Steier, L.; Ahmad, S.; Luterbacher, J. S.; Mayer, M. T.; Luo, J. S.; Grätzel, M., Solar conversion of CO2 to CO using Earth-abundant electrocatalysts prepared by atomic layer modification of CuO. *Nat. Energy* **2017,** *2* (7), 17087.
14. Wang, S. B.; Guan, B. Y.; Lu, Y.; Lou, X. W., Formation of hierarchical InS-CdInS heterostructured nanotubes for efficient and stable visible light CO reduction. *J. Am. Chem. Soc.* **2017,** *139* (48), 17305-17308.
15. Li, Y. G.; Wang, H. L.; Xie, L. M.; Liang, Y. Y.; Hong, G. S.; Dai, H. J., MoS nanoparticles grown on graphene: An advanced catalyst for the hydrogen evolution reaction. *J. Am. Chem. Soc.* **2011,** *133* (19), 7296-7299.
16. Subbaraman, R.; Tripkovic, D.; Strmcnik, D.; Chang, K.-C.; Uchimura, M.; Paulikas, A. P.; Stamenkovic, V.; Markovic, N. M., Enhancing hydrogen evolution activity in water splitting by tailoring Li+-Ni(OH)2-Pt interfaces. *Science* **2011,** *334* (6060), 1256-1260.

17. Gunjakar, J. L.; Kim, T. W.; Kim, H. N.; Kim, I. Y.; Hwang, S. J., Mesoporous layer-by-layer ordered nanohybrids of layered double hydroxide and layered metal oxide: Highly active visible light photocatalysts with improved chemical stability. *J. Am. Chem. Soc.* **2011,** *133* (38), 14998-15007.

18. Yang, M. Q.; Xu, Y. J.; Lu, W. H.; Zeng, K. Y.; Zhu, H.; Xu, Q. H.; Ho, G. W., Self-surface charge exfoliation and electrostatically coordinated 2D hetero-layered hybrids. *Nat. Commun.* **2017,** *8* (1), 14224.

19. Ma, W.; Ma, R. Z.; Wang, C. X.; Liang, J. B.; Liu, X. H.; Zhou, K. C.; Sasaki, T., A superlattice of alternately stacked Ni-Fe hydroxide nanosheets and graphene for efficient splitting of water. *Acs Nano* **2015,** *9* (2), 1977-1984.

20. Jiang, L. L.; Duan, J. J.; Zhu, J. W.; Chen, S.; Antonietti, M., Iron-cluster-directed synthesis of 2D/2D Fe-N-C/MXene superlattice-like heterostructure with enhanced oxygen reduction electrocatalysis. *Acs Nano* **2020,** *14* (2), 2436-2444.

21. Cai, X. K.; Ozawa, T. C.; Funatsu, A.; Ma, R. Z.; Ebina, Y.; Sasaki, T., Tuning the surface charge of 2D oxide nanosheets and the bulk-scale production of superlatticelike composites. *J. Am. Chem. Soc.* **2015,** *137* (8), 2844-2847.

22. Naguib, M.; Kurtoglu, M.; Presser, V.; Lu, J.; Niu, J. J.; Heon, M.; Hultman, L.; Gogotsi, Y.; Barsoum, M. W., Two-dimensional nanocrystals produced by exfoliation of TiAlC. *Adv. Mater.* **2011,** *23* (37), 4248-4253.

23. VahidMohammadi, A.; Rosen, J.; Gogotsi, Y., The world of two-dimensional carbides and nitrides (MXenes). *Science* **2021,** *372* (6547), eabf1581.

24. Naguib, M.; Mashtalir, O.; Carle, J.; Presser, V.; Lu, J.; Hultman, L.; Gogotsi, Y.; Barsoum, M. W., Two-dimensional transition metal carbides. *ACS Nano* **2012,** *6* (2), 1322-1331.

25. Anasori, B.; Lukatskaya, M. R.; Gogotsi, Y., 2D metal carbides and nitrides (MXenes) for energy storage. *Nat. Rev. Mater.* **2017,** *2* (2), 16098.

26. Smith, A. J.; Trimm, D. L., The Preparation of Skeletal Catalysts. *Annual Review of Materials Research* **2005,** *35* (1), 127-142.

27. Lee, K.; Kim, S. W.; Toda, Y.; Matsuishi, S.; Hosono, H., Dicalcium nitride as a two-dimensional electride with an anionic electron layer. *Nature* **2013,** *494* (7437), 336-340.

28. Zhang, X.; Xiao, Z. W.; Lei, H. C.; Toda, Y.; Matsuishi, S.; Kamiya, T.; Ueda, S.; Hosono, H., Two-dimensional transition-metal electride YC. *Chem. Mater.* **2014,** *26* (22), 6638-6643.

29. McRae, L. M.; Radomsky, R. C.; Pawlik, J. T.; Dru, D. L.; Sundberg, J. D.; Lanetti, M. G.; Donley, C. L.; White, K. L.; Warren, S. C., ScC, a 2D semiconducting electride. *J. Am. Chem. Soc.* **2022,** *144* (24), 10862-10869.

30. Druffel, D. L.; Kuntz, K. L.; Woomer, A. H.; Alcorn, F. M.; Hu, J.; Donley, C. L.; Warren, S. C., Experimental demonstration of an electride as a 2D material. *J. Am. Chem. Soc.* **2016,** *138* (49), 16089-16094.

31. Bae, S.; Espinosa-García, W.; Kang, Y. G.; Egawa, N.; Lee, J.; Kuwahata, K.; Khazaei, M.; Ohno, K.; Kim, Y. H.; Han, M. J.; Hosono, H.; Dalpian, G.

M.; Raebiger, H., MXene phase with C structure unit: A family of 2D electrides. *Adv. Funct. Mater.* **2021,** *31* (24), 2100009.
32. Cai, W.; Zhou, C.; Hu, X.; Jiao, T.; Liu, Y.; Li, L.; Li, J.; Kitano, M.; Hosono, H.; Wu, J., Quasi-two-dimensional intermetallic electride CeRuSi for efficient alkaline hydrogen evolution. *ACS Catal.* **2023,** *13* (7), 4752-4759.
33. Shen, S. J.; Hu, Z. Y.; Zhang, H. H.; Song, K.; Wang, Z. P.; Lin, Z. P.; Zhang, Q. H.; Gu, L.; Zhong, W. W., Highly active Si sites enabled by negative valent Ru for electrocatalytic hydrogen evolution in LaRuSi. *Angew Chem Int Edit* **2022,** *61* (32), e202206460.
34. Wu, J. Z.; Li, J.; Gong, Y. T.; Kitano, M.; Inoshita, T.; Hosono, H., Intermetallic electride catalyst as a platform for ammonia synthesis. *Angew Chem Int Edit* **2019,** *58* (3), 825-829.
35. Li, J.; Wu, J. Z.; Wang, H. Y.; Lu, Y. F.; Ye, T. N.; Sasase, M.; Wu, X. J.; Kitano, M.; Inoshita, T.; Hosono, H., Acid-durable electride with layered ruthenium for ammonia synthesis: boosting the activity selective etching. *Chem. Sci.* **2019,** *10* (22), 5712-5718.
36. Wu, J. Z.; Lu, E. D.; Li, J.; Lu, Y. F.; Kitano, M.; Fredrickson, D. C.; Inoshita, T.; Hosono, H., Pseudogap control of physical and chemical properties in CeFeSi-type intermetallics. *Inorg. Chem.* **2019,** *58* (4), 2848-2855.
37. Beche, E.; Peraudeau, G.; Flaud, V.; Perarnau, D., An XPS investigation of (La2O3)1-x(CeO2)2x(ZrO2)2 compounds. *Surf. Interface Anal.* **2012,** *44* (8), 1045-1050.
38. Jensen, D. S.; Kanyal, S. S.; Madaan, N.; Vail, M. A.; Dadson, A. E.; Engelhard, M. H.; Linford, M. R., Silicon (100)/SiO2 by XPS. *Surf. Sci. Spectra* **2013,** *20* (1), 36-42.
39. Dang, Y. L.; Wu, T. L.; Tan, H. Y.; Wang, J. L.; Cui, C.; Kerns, P.; Zhao, W.; Posada, L.; Wen, L. Y.; Suib, S. L., Partially reduced Ru/RuO composites as efficient and pH-universal electrocatalysts for hydrogen evolution. *Energy Environ. Sci.* **2021,** *14* (10), 5433-5443.
40. Shah, K.; Dai, R.; Mateen, M.; Hassan, Z.; Zhuang, Z.; Liu, C.; Israr, M.; Cheong, W. C.; Hu, B.; Tu, R.; Zhang, C.; Chen, X.; Peng, Q.; Chen, C.; Li, Y., Cobalt single atom incorporated in ruthenium oxide sphere: A robust bifunctional electrocatalyst for HER and OER. *Angew. Chem. Int. Ed.* **2021,** *61* (4), e202114951.
41. Shen, J. Y.; Adnot, A.; Kaliaguine, S., An ECSA study of the interaction of oxygen with the surface of ruthenium. *Appl. Surf. Sci.* **1991,** *51* (1-2), 47-60.
42. Momma, K.; Izumi, F., for three-dimensional visualization of crystal, volumetric and morphology data. *J. Appl. Crystallogr.* **2011,** *44* (6), 1272-1276.
43. Dinh, C. T.; Jain, A.; de Arquer, F. P. G.; De Luna, P.; Li, J.; Wang, N.; Zheng, X. L.; Cai, J.; Gregory, B. Z.; Voznyy, O.; Zhang, B.; Liu, M.; Sinton, D.; Crumlin, E. J.; Sargent, E. H., Multi-site electrocatalysts for hydrogen evolution in neutral media by destabilization of water molecules. *Nat. Energy* **2019,** *4* (2), 107-114.
44. Zhu, Z. J.; Yin, H. J.; He, C. T.; Al-Mamun, M.; Liu, P. R.; Jiang, L. X.;

Zhao, Y.; Wang, Y.; Yang, H. G.; Tang, Z. Y.; Wang, D.; Chen, X. M.; Zhao, H. J., Ultrathin transition metal dichalcogenide/3d metal hydroxide hybridized nanosheets to enhance hydrogen evolution activity. *Adv. Mater.* **2018,** *30* (28), 1801171.
45. Liu, Z. J.; Qi, J.; Liu, M. X.; Zhang, S. M.; Fan, Q. K.; Liu, H. P.; Liu, K.; Zheng, H. Q.; Yin, Y. D.; Gao, C. B., Aqueous synthesis of ultrathin platinum/non-noble metal alloy nanowires for enhanced hydrogen evolution activity. *Angew Chem Int Edit* **2018,** *57* (36), 11678-11682.
46. Kresse, G.; Joubert, D., From ultrasoft pseudopotentials to the projector augmented-wave method. *Phys. Rev. B* **1999,** *59* (3), 1758-1775.
47. Kresse, G.; Furthmuller, J., Efficient iterative schemes for ab initio total-energy calculations using a plane-wave basis set. *Phys. Rev. B* **1996,** *54* (16), 11169-11186.
48. Perdew, J. P.; Burke, K.; Ernzerhof, M., Generalized gradient approximation made simple. *Phys. Rev. Lett.* **1996,** *77* (18), 3865-3868.
49. Korzhavyi, P. A.; Abrikosov, I. A.; Johansson, B.; Ruban, A. V.; Skriver, H. L., First-principles calculations of the vacancy formation energy in transition and noble metals. *Phys. Rev. B* **1999,** *59* (18), 11693-11703.
50. Grimme, S.; Antony, J.; Ehrlich, S.; Krieg, H., A consistent and accurate ab initio parametrization of density functional dispersion correction (DFT-D) for the 94 elements H-Pu. *J. Chem. Phys.* **2010,** *132* (15), 154104.
51. Henkelman, G.; Uberuaga, B. P.; Jónsson, H., A climbing image nudged elastic band method for finding saddle points and minimum energy paths. *J. Chem. Phys.* **2000,** *113* (22), 9901-9904.
52. Henkelman, G.; Jónsson, H., A dimer method for finding saddle points on high dimensional potential surfaces using only first derivatives. *J. Chem. Phys.* **1999,** *111* (15), 7010-7022.
53. Wang, V.; Xu, N.; Liu, J. C.; Tang, G.; Geng, W. T., VASPKIT: A user-friendly interface facilitating high-throughput computing and analysis using VASP code. *Comput. Phys. Commun.* **2021,** *267*, 108033.

**Acknowledgments:**

This work is supported by Guangdong Basic and Applied Basic Research Foundation (2022A1515011295), National Natural Science Foundation of China (No. 52173284), Shenzhen fundamental research funding (JCYJ20210324115809026), Shenzhen Key Laboratory Program (ZDSYS20210709112802010), and Guangdong Provincial Department of Education Innovation Team Program (2021KCXTD012). H.H. acknowledges the funding from the JSPS Kakenhi Grants-in-Aid (17H06153) and JST-Mirai Program (JPMJMI21E9). We thank Hitoshi Abe and Yasuhiro Niwa for their support in the XAFS experiments.

**Author contributions:**

J.W. and W.C. proposed the idea and designed the project. J.W. and H.H. supervised the research. J.W., W.C., X. Hu, M.S., M.K., and T.Y. conducted the materials synthesis and characterization. W.C. and C.L. performed the electrochemical experiments, and analyzed the experimental data. X. He and T.K. performed the DFT calculations. W.C., X. He, J.W. and H.H. co-wrote the paper. All authors discussed the results and commented on the manuscript.

**Competing interests:** Authors declare that they have no competing interests.

**Data and materials availability:** All data are available in the manuscript, the Supporting information.